\documentstyle[preprint,aps]{revtex}
\begin{document}
\draft
\tighten
\title{INSTANTON SYMMETRIES \\ AND \\ 
    THE ENTROPY OF COMPACT MANIFOLDS}

\author{Marika Taylor-Robinson \footnote{E-mail:
    mmt14@damtp.cam.ac.uk}}
\address{Department of Applied Mathematics and Theoretical Physics,
\\University of Cambridge, Silver St., Cambridge. CB3 9EW}
\date{\today}
\maketitle

\begin{abstract}
{Many Euclidean Einstein manifolds possess continuous symmetry groups
  of at least one parameter and we consider here a classification
scheme of $d$ dimensional compact manifolds based on the existence of
  such a one parameter group in terms of the fixed point sets of the
isometries. We discuss applications of such a classification scheme,
  including the geometric interpretation of the entropy; 
there are intrinsic contributions to the entropy from the volumes of
$(d-2)$ dimensional fixed point sets and contributions related to
the cohomology structure of the orbit space of the isometry. 
We consider the relevance of such a
decomposition of the entropy in the context of the no boundary
proposal and cosmological processes, and generalise the discussion to 
compact solutions of gravity coupled to scalar and gauge fields.}
\end{abstract}
\narrowtext

\section{Introduction}
\noindent

Euclidean Einstein manifolds arise as instanton solutions of the classical
Euclidean field equations of not only $d$ dimensional gravity
but also of supergravity theories, 
with a constant dilaton and all other fields except the graviton
vanishing. Many instanton solutions possess continuous symmetry groups of at
least one parameter; indeed in many cases we consider dimensional reduction of
$d$ dimensional solutions to $(d-1)$ dimensions along closed orbits 
of circle isometries. We consider here a classification scheme
of $d$ dimensional Euclidean Einstein manifolds
based on the existence of such a one parameter group, in terms of
the fixed point sets of the isometries, generalised nuts and bolts. 
This is a generalisation of the
four-dimensional case analysed in \cite{GH}; the action of fixed point
sets of isometries has also been considered
in \cite{GPP}, \cite{Do2} and \cite{Ra}. 

Such a classification scheme has various applications, perhaps the
most important of which is the geometric interpretation of the
entropy. It is well known that black holes have
an intrinsic entropy proportional to one quarter of the volume of the
horizon. In addition cosmological event horizons have an associated
entropy equal to one quarter of their volume; 
this entropy can be derived using the Euclidean path integral
approach. In terms of the nuts and bolt terminology, the
$(d-2)$ dimensional horizon is a fixed point set of the imaginary time
Killing vector, and contributes an entropy proportional to its proper
volume. Recently the advent of D-brane technology in string theory 
has permitted a microscopic derivation of black hole entropy for
particular classes of near BPS states. 

However, for four dimensional compact solutions it is known that
there are contributions to the gravitational entropy
not only from the areas of the bolts but also from the nut charges of
the nuts and bolts. In \cite{GH} the entropy of a four
dimensional compact Einstein manifold with no boundary 
admitting at least a circle subgroup was found to be
\begin{equation}
S = \sum_{bolts} \frac{V_{2}}{4 G_{4}} + \sum_{nuts, bolts}
\frac{\beta}{16\pi G_{4}} \Psi_{a} \int_{M_{a}^{2}} F, \label{4dent}
\end{equation}
where the nuts and bolts are zero and two dimensional fixed point sets
respectively. $\Psi_{a}$ is a scalar potential evaluated at the $a$th
fixed point set, and $F$ is the Kaluza-Klein two form gauge field 
obtained upon dimensional reduction along closed orbits of the
isometry. Further discussions of the r\^{o}le of the nut charge were
presented recently in \cite{cjh_1}.

The main object of our classification scheme is to
extend this geometric interpretation of the entropy in terms of fixed point
sets to general dimensions. What we find is that
$(d-2)$ dimensional bolts have an intrinsic entropy related to their
volume. There are additional contributions to the entropy from 
all bolts of lower dimension and $(d-2)$ dimensional bolts which have
non-trivial normal bundles; these contributions can be thought of as the
generalisations of the four-dimensional nut potential terms.

Although in four dimensions one can represent the nut contributions 
to the entropy in terms the properties of the fixed point sets only,
in higher dimensions the situation is considerably more
complicated. Non trivial $(d-3)$ cohomology of the $(d-1)$ dimensional
orbit space plays a role as does the nut type behaviour of individual
fixed point sets. The total contribution to the entropy can best be
represented as 
\begin{equation}
S = \sum_{a} \frac{V_{a}}{4G_{d}} + 
\frac{\beta}{16 \pi G_{d}} \int_{\Sigma} F \wedge \bar{G},
\end{equation}
where $\bar{G}$ is a $(d-3)$ form related to the dual of $F$ in
the orbit space $\Sigma$, and the integral is taken over this space. 
In general both $\bar{G}$ and $F$ have non-zero periods and 
there is no natural way to split the integral into individual 
contributions from fixed point sets and Dirac string type
singularities within the orbit space. 

\bigskip 

This decomposition of the Euclidean action, and thence the entropy, of
compact manifolds in terms of the action of isometries has
implications for cosmological processes. In the context of the no
boundary proposal, we can express the entropy of
the Lorentzian solution in terms of the action of the isometry on the
original Euclidean manifold. Thence we are able to demonstrate
explicitly that cosmological pair creation of black branes is
associated with an entropy dependent on the volumes of the horizons,
and that monopole pair creation within an expanding background is 
associated with an entropy dependent on the monopole charges. 

Having discussed the action of isometries on Euclidean Einstein
manifolds, it is natural to consider the extensions to Euclidean
solutions of gravity coupled to scalar and gauge fields. We
find that the same decomposition of the action holds, but if the 
``electric'' part of the gauge field is non vanishing, there is an
additional term in the action dependent on this part of the field. 
In this context, ``electric'' means that if we consider the action 
of an isometry $\partial_{\tau}$ the 
${\mathcal H}_{\tau...}$ components of the gauge 
field are non zero. If we analytically continue the solution and
$\tau$ is interpreted as an imaginary time coordinate, this part of
the gauge field will indeed be electric. 

In the context of the no
boundary proposal, we find that the additional term in the action
can be removed by imposing a constraint on the nucleation surface;
this constraint has the physical interpretation of fixing the charge on
the hypersurface. We can then show that cosmological pair creation of
generic charged black dilatonic branes is associated with an entropy dependent
on the horizon volumes. The treatment of extreme black
holes within this formalism requires a more careful treatment of the
boundary terms, and will be considered briefly.

\bigskip

The plan of this paper is as follows. In \S\ref{pos} we give a
brief discussion of the properties of higher dimensional symmetry
groups. In \S\ref{acs} we discuss the decomposition of the action of
compact solutions in terms of the properties of the fixed point sets
of the action of a circle isometry group. In \S\ref{pca} we consider
further the contributions to the action from the non-trivial cohomology of
$\Sigma$. We discuss the entropy of such
solutions, and the cosmological relevance in \S\ref{eci}, and consider
the generalisation to gauge field theories in \S\ref{ggft}. 
In \S\ref{exsol} we discuss the inclusion of boundaries to the compact
manifold, and the applications to extreme solutions. 

\section{Properties of symmetries} \label{pos}
\noindent

We will be considering solutions of the Euclidean action of
$d$ dimensional Einstein gravity
(omitting cosmological constant and boundary terms for the meantime)
\begin{equation}
{S_{E} = - \frac{1}{16\pi G_{d}} \int_{M} d^dx \sqrt{\hat{g}} R_{d}},
\label{daction}
\end{equation}
where $\hat{g}$ is the determinant of the $d$ dimensional metric
and $R_{d}$ is the Ricci scalar. $G_{d}$ is the $d$ dimensional newton
constant. The $d$ dimensional oriented manifold $M$
will in general have a $(d-1)$ dimensional boundary at infinity 
which we denote as $\partial M$. 

Many solutions of interest admit continuous symmetry groups of at
least two parameters, and we assume here the existence of at least 
a one parameter group.
A solution admitting a Killing vector $k$ with closed orbits can be
written in terms of $(d-1)$ dimensional fields, which we refer to as
the dilaton $\phi$, gauge potential $A_{i}$ and metric $g_{ij}$, as
\begin{equation}
{ds^2 = e^{\frac{-4\phi}{\sqrt{d-2}}}(dx^d + A_{i}dx^i)^2 +
e^{\frac{4\phi}{(d-3)\sqrt{d-2}}}g_{ij}dx^idx^j},
\label{line-el}
\end{equation}
where we take the Killing vector to be $\partial_{x^d}$ of period
$\beta = 2 \pi \mu$. In the context of Kaluza-Klein theories it is
perhaps more conventional to let $A_{i} \equiv 2 A'_{i}$;
we choose the normalisation here for 
later convenience when we compare the four dimensional limit of our
results with those obtained in \cite{GH}. 
The action can be expressed in terms of the lower dimensional fields as 
\begin{equation}
S_{E} = - \frac{1}{16\pi G_{d-1}} \int_{\Sigma} d^{d-1}x \sqrt{g} \left[ R -
\frac{4}{d-3}(\partial\phi)^2 - 
\frac{1}{4}e^{\frac{-4\sqrt{d-2}}{d-3} \phi}F^2 \right],
\label{xxy}
\end{equation}
where $G_{d} = \beta G_{d-1}$. We refer to 
the $(d-1)$ dimensional manifold we obtain after dividing out by the 
$U(1)$ isometry as $\Sigma$ with $(d-2)$ dimensional boundary
$\partial \Sigma$. This is precisely the dimensional reduction procedure that
is used in Kaluza-Klein theories, hence our notation for the $(d-1)$
dimensional fields.

\bigskip

If the isometry generated by the Killing vector has fixed point sets,
then the metric $g_{ij}$ will be singular at these points. Denote by
$\mu_{\tau} : M \rightarrow M$ the action of the group where $\tau$ is
the group parameter. At a fixed point, the action of $\mu_{\tau}$ on
the manifold $M$ gives rise to an isometry $\mu_{\tau}^{*} : T_{p}(M)
\rightarrow T_{p}(M)$ where $\mu_{\tau}^{*}$ is generated by the
antisymmetric tensor $k_{M;N}$. Vectors in the kernel $V$ of $k_{M;N}$
leave directions in the tangent space at a fixed point invariant under
the action of the symmetry. The image of the invariant subspace of
$T_{p}(M)$ under the exponential map will not be moved by
$\mu_{\tau}$, and so will constitute a submanifold of fixed points of
dimension $p$ where $p$ is the dimension of the kernel of $V$. Since
the rank of an antisymmetric matrix must be even, the dimension of the
invariant subspace may take the values $0,2,..,d$ for $d$ even, and
$0,.,(d-1)$ for $d$ odd. If the fixed point set is decomposed into
connected components, each connected component is a closed 
totally geodesic submanifold of even codimension \cite{Be}.

\bigskip

In four dimensions, the possible fixed point sets are 2-dimensional
submanifolds (bolts) and fixed points (nuts). In higher dimensions,
generalised bolts and nuts are possible. There are at most $[d/2]$
eigenvalues of $k_{M;N}$, $\lbrace n_{i} \rbrace$, where $[n]$ denotes
the integer part of
$n$. If the eigenvalues are rationally related, the action of
$\mu_{\tau}$ will be periodic, with the integers relating the
eigenvalues determining the number of rotations in distinct orthogonal
planes in $T_{p}(M)$ induced by one orbit of the isometry. If one pair
of eigenvalues are not rationally related, the orbits of a vector in
$T_p(M)$ under the action of $\mu_{\tau}^*$ is dense in the torus $C$
consisting of all vectors of the form 
$\mu_{\tau_2}^{1*} \cdot \mu_{\tau_2}^{2*}$ where
$\mu_{\tau}^{*}$ has rank $n$,
$\mu_{\tau_{1}}^{1*}$ has rank $(n-1)$ and  
$\mu_{\tau}^{*} = \mu_{\tau_1}^{1*} \cdot \mu_{\tau_2}^{2*}$. 
All scalar invariants must be constant over each
torus in $M$ of the form $\exp(C(M))$ for each $X \in T_{p}(M)$, and
since scalar invariants characterise the metric it follows that 
$\mu_{\tau_1}^{1*}$ and $\mu_{\tau_2}^{2*}$ must actually correspond
to independent isometries of the metric on M. One can then take
appropriate combinations of the Killing vectors such that the orbits
are periodic; we thus assume that the action of the isometry group is
periodic.

\bigskip

It is useful to express this action of the group on the fixed point set
in the following way. Let G be a finite or closed Lie group acting 
on the oriented manifold $M$.  
We consider the action of an element $g \in G$; we denote fixed points
of the action of the isometry as 
$M^{g} = \lbrace x: gx = x \rbrace$, and construct the normal
bundle $N^{g}_{x}$ over $X^{g}$. Then in the neighbourhood of the 
fixed point set the action of $g$ on the normal bundle is effective and
$N^{g}$ may be equivariantly identified with a neighbourhood of 
$M^{g}$ in $M$:
\begin{equation}
(z_{1},..,z_{s}) \leftrightarrow (z_{1},..,z_{s},x_{1},..,x_{d-2s}),
\end{equation}
where we choose coordinates on the fixed point set such that:
\begin{equation}
(x_{1},..,x_{d}) = (x_{1},..,x_{d-2s},0,..0),
\end{equation}
and $z_{i} \equiv \rho_{i} e^{\psi_{i}}$ are 
{\it{complex}} coordinates in a neighbourhood of 
$M^{g}$, which are acted on by $g$ as:
\begin{equation}
g(z_{1},..,z_{s}) = (e^{i n_{1}\theta}z_{1},..,e^{i n_{s}\theta}z_{s}),
\end{equation}
where $\theta$ is the group parameter and takes values between $0$
and $2 \pi$. That is, locally we can decompose the normal bundle as a
direct sum of complex line bundles.
Expressed in terms of the metric, in a small neighbourhood of a
fixed point set of dimension $d-2k$, we can write the metric as
\begin{equation}
ds^2 = \sum^{k} (d\rho_{i}^2 + \rho_{i}^{2} d\psi_{i}^{2}) +
d\bar{s}_{d-2k}^2,
\end{equation}
where each $\psi_{i}$ has period $2\pi$, and the Killing vector
is $\partial_{\theta} = \sum_{i} n_{i} \partial_{\psi_{i}}$. We shall
find this form of the metric to be useful in the following sections. 

\bigskip

Where the symmetry group is more than one-dimensional, different 
choices of the one parameter subgroup may lead to different numbers
and locations of nuts and bolts. However topological invariants of
the manifold - the Euler characteristic and the Hirzebruch signature - are 
evidently independent of the choice of circle subgroup. By the
Lefschetz fixed point theorem, the Euler characteristic for a compact 
manifold without boundary may be decomposed (see for example
\cite{EGH}) as:
\begin{equation}
\chi(M) = \sum_{i} \chi(M_{i}^{g}), \label{eul}
\end{equation}
where we sum over the fixed point sets, and the Euler characteristic
of a point is one.

In four dimensions, the G-signature theorem takes the particularly simple
form
\begin{equation}
\tau = \sum_{nuts} -\cot \frac{p\theta}{2} \cot \frac{q\theta}{2} 
+ \sum_{bolts} \cal{Y} \rm{cosec}^2 \frac{\theta}{2}, \label{sig4}
\end{equation}
where $\cal{Y}$ is the self-intersection number of a bolt, and the
integers $p,q$ characterise the normal bundle over the nut fixed point
set. Expanding in powers of $\theta$, one then obtains constraints on
the nut and bolt parameters given in \cite{GH}. 

However, in general dimensions, the G-signature theorems take a much
more complicated form, and we will not use them here. We mention only
the most simple case of nut fixed point sets, assuming that $d$ is
even. We may then express the signature as \cite{HZ}:
\begin{equation}
\tau = \sum_{nuts} \prod_{i=1}^{d/2} (-i \cot \frac{n_{i}\theta}{2}).
\end{equation}
This is imaginary when the dimension is not a multiple
of four, whereas the dimensions of the cohomology groups are real, and
the signature vanishes. For fixed point sets of general dimension, we
would expect the signature theorem to include for example terms involving the
signature of the self-intersection manifold of the fixed point set.
Schematically the Euler characteristic is a weighted sum over
the number of fixed point sets, whilst the signature depends on the
structure of the normal bundle over the fixed point set. 

In fact, the form of the signature theorem is an indication that the
analysis of higher dimensional Einstein manifolds in terms of fixed
point sets is much more complicated than in four dimensions. As we
shall see, many simplifications of the analysis in \cite{GH} 
arise from the existence of a type of electromagnetic duality in four
dimensions. 

\bigskip

We briefly mention here examples of complete non-singular Einstein
manifolds which are of interest physically. In order for a compact 
Einstein space $M$ to admit continuous
isometries, the Ricci scalar must be non-negative and, if one excludes the
case where $M$ contains flat circle factors, $M$ can admit Killing
vectors only if the Ricci scalar is strictly positive. 

The obvious examples of positive curvature compact Einstein manifolds
are homogeneous manifolds, $G/H$, where $G$ is the isometry group and
$H$ is the isotropy group, which admit Einstein metrics. The simplest
example is the $d$ dimensional sphere, with canonical metric, 
which may be viewed as the homogeneous manifold $SO(d+1)/O(d)$. 
The $SO(d+1)$ isometry group is generated
by $(d+1)$-dimensional anti-symmetric matrices of rank $0, 2, ...$.
In the case of rank 2, there is a $(d-1)$-plane through the origin which
is not moved by the rotation, and the intersection of this with the
$d$-sphere is a $(d-2)$-dimensional sphere. Higher rank matrices leave
smaller spheres invariant. Another interesting 
example is complex projective space of complex dimension
$n$ which may be viewed as the homogeneous manifold
$SU(n+1)/S(U(1)U(n))$; we will discuss this further in \S\ref{pca}.

Many of the manifolds discussed \cite{GH} have natural extensions 
to higher dimensions. For example, one may take the metric product of
a two sphere with a $(d-2)$ sphere; this is a (regular) limit of the
Schwarzschild de Sitter solution and we shall consider it further in
\S\ref{eci}. Furthermore, one could consider
inhomogeneous Einstein manifolds such as those constructed in
\cite{PP} and \cite{GPP}, although we shall not do so here. 

\section{Action of compact solutions} \label{acs}
\noindent

Given an Euclidean Einstein manifold, we are interested in calculating
its action, since this is important in describing the thermodynamics
of the system, and gives a measure of the probability for a decay
into the instanton to occur. 
In this section we will rewrite the action in terms of the lower
dimensional fields and a $(d-3)$ form which we will define. 
For compact manifolds, we can then
obtain an expression for the action entirely in terms of 
characteristics of the orbit space of the isometry. 

The action of a circle subgroup of the
isometry group of $M$ defines a fibering $\pi : M-C \rightarrow B$, where $C$
is the fixed point set of the isometry and $B$ is a
$d$ dimensional space of non-trivial orbits. The metric on $M$ can
be expressed in the form (\ref{line-el}), with the gauge field $A_{i}$
invariant under the gauge transformation $A_{i}' = A_{i} - \partial_i a$
and the gauge invariant field strength being $F_{ij} = \partial_i A_j -
\partial_j A_i$.
The Bianchi identity implies that:
\begin{equation}
D_{[i}F_{jk]} = 0, 
\end{equation}
or in form notation $dF=0$, where we define the covariant derivative
with respect to the metric $g_{ij}$. The equation of motion for 
$F_{ij}$ derived from the Lagrangian (\ref{xxy}) is
\begin{equation}
D_{i}(e^{\frac{-4\sqrt{d-2}\phi}{d-3}} F^{ij}) = 0,
\label{eqmot}
\end{equation}
which may be rewritten in the form $D_{i}F^{ij} = J^{j}$ where $J^{j}$
is the conserved current
\begin{equation}
J^{j} = \frac{4\sqrt{d-2}}{d-3} (\partial_{i}\phi)F^{ij}. \label{curr}
\end{equation}
In form notation, we may express this as 
$dG = *J$, where $G_{i_{1}...i_{d-3}}$ is the dual
field strength defined by
\begin{equation}
G^{i_{1}...i_{d-3}} = \frac{1}{2 \sqrt{g}}
  \epsilon^{i_{1}...i_{d-1}} F_{i_{d-2}i_{d-1}},
\end{equation}
with $\epsilon_{1...(d-1)} = 1$. The Bianchi identity for the 
field $F$ dualises to give the field equation for $G$, $d*G = 0$. 

For clarity we mention here that the action
may also be dualised by making the transformations 
\begin{equation}
\bar\phi = -\phi, \hspace{5mm}
\tilde{F} = e^{\frac{-4\sqrt{d-2}\phi}{d-3}} \ast F.
\end{equation}
It is this duality which we commonly use in supergravity theories;
it exchanges the field equations and the Bianchi identities. In the
absence of the cosmological term, the equations of motion from the
resultant action admit solutions in which the metric is unchanged from
the corresponding solution in the original theory but
``electric'' fields are exchanged for ``magnetic'' fields. In the
presence of the cosmological term, solutions of the equations of motion
derived from the dualised action are not solutions of the original
equations of motion. The ``duality'' we use here simply re-expresses
the original solution in terms of different fields. 
 
Associated with the conserved current (\ref{curr}), there is a conserved
(``electric'') charge:  
\begin{equation}
Q_{e} = \int_{M^{(d-2)}} J_i d\sigma^i,
\end{equation}
or in form notation $Q_{e}= \int_{M^{d-2}} *J = \int_{M^{d-2}} dG$. 
It is important to
note here that in general dimensions there is no 
such conserved quantity associated with the dual field strength;
there is no ``magnetic'' charge. In four dimensions, one can define a
conserved charge by $P_{m} = \int_{M^2} F$. The Bianchi identity
implies that the total charge vanishes for compact solutions, although
we may define non-zero charges within closed two-dimensional submanifolds. 
Expressed in the language of \cite{GH}, the nut charges associated
with individual fixed point sets sum to zero for a compact manifold,
which has a simple interpretation in terms of the G-signature 
theorem (\ref{sig4}).  

For higher-dimensional solutions, one cannot define a unique
two sphere at infinity and there is no ``magnetic'' charge. So as we
stated earlier ``electromagnetic'' duality is a concept confined to
four dimensions. However, as we shall see, there is a
straightforward generalisation of the decomposition of the action in
terms of the properties of the orbit space of the isometry. 

\bigskip

The $d$ dimensional Euclidean action, including boundary and
cosmological constant terms, is
\begin{equation}
S_{E} = - \frac{1}{16\pi G_{d}} \int_{M} d^{d}x \sqrt{\hat{g}} (R_{d} - m) -
\frac{1}{8 \pi G_{d}} \int_{\partial M} d^{d-1}x \sqrt{b} 
({\cal{K}} - {\cal{K}}_{0}),
\end{equation}
where $b$ is the induced metric on the boundary and $\cal{K}$ is the trace
of the second fundamental form in the $d$ dimensional metric defined
with respect to a suitable background geometry ${\cal{K}}_{0}$. 

We choose the cosmological constant term such the 
solution is Einstein with $R_{MN} = \Lambda\
\hat{g}_{MN}$, which implies that $m = (d-2)\Lambda$. 
After dimensional reduction along a closed orbit of the isometry 
the volume term in the action, in the Einstein frame, becomes 
\begin{equation}
S_{E} = - \frac{1}{16\pi G_{d-1}} \int_{\Sigma} d^{d-1}x \sqrt{g} [R -
\frac{4}{d-3}(\partial\phi)^2 
 - m e^{\frac{4\phi}{\sqrt{d-2}(d-3)}}
- \frac{1}{4}e^{\frac{-4\sqrt{d-2}}{d-3}\phi} F^2]. \label{orac}
\end{equation}
We can express the action in terms of the dual field strength 
$G = \ast F$, with the appropriate field equations being:
\begin{eqnarray}
D_{i_{1}} G^{i_{1}...i_{d-3}} = 0; \\
D_{[i}(e^{-\frac{4\sqrt{d-2}\phi}{d-3}}G_{i_{1}...i_{d-3}]}) = 0,
\nonumber 
\end{eqnarray}
which are equivalent to those given previously, but expressed in
coordinate form. To obtain these field equations from a dualised
action, we require that the action is stationary under variations of
the fields subject to the constraint that the dual field strength is
conserved; 
we thus define the dualised action a constraint term to the action
\begin{eqnarray}
\bar S_{E} = -\frac{1}{16\pi G_{d-1}} \int_{\Sigma} 
d^{d-1}x \sqrt{g} \lbrace R - m
e^{\frac{4\phi}{\sqrt{d-2}(d-3)}} - \frac{4}{d-3}(\partial\phi)^2
\nonumber \\
- \frac{1}{2(d-3)!} e^{\frac{-4\sqrt{d-2}}{d-3}\phi} G^2 +
\frac{1}{(d-4)!}
G^{i_{1}....i_{d-3}}  D_{[i_{1}}\Psi_{i_{2}...i_{d-3}]} \rbrace. 
\label{feq}
\end{eqnarray}
Then the field equation for $\Psi$ gives the constraint equation for the 
$(d-3)$-form $G$, whilst 
variation of $G$ gives the defining equation for the potential $\Psi$:
\begin{equation}
e^{\frac{-4 \sqrt{d-2}\phi}{d-3}}
G_{i_{1}...i_{d-3}} = (d-3) D_{[i_{1}}\Psi_{i_{2}...i_{d-3}]}.   
\end{equation}
We then rewrite the action in terms of the potential as
\begin{eqnarray}
\bar S_{E} = - \frac{1}{16 \pi G_{d-1}} \int_{\Sigma} 
d^{d-1}x \sqrt{g} \lbrace R -
m e^ {\frac{4\phi}{\sqrt{d-2}(d-3)}}  - \frac{4}{d-3}
(\partial\phi)^{2} \\
 + \frac{(d-3)}{2(d-4)!} e^{\frac{4\sqrt{d-2}}{d-3}\phi}
(D_{[i_{1}}\Psi_{i_{2}...i_{d-3}]})^2 \rbrace. \nonumber 
\end{eqnarray} 
The field equations may be expressed as:
\begin{eqnarray}
D_{i_{1}} ( e^{\frac{4\sqrt{d-2}\phi}{d-3}} D^{[i_{1}} 
  \Psi^{i_{2}...i_{d-3}]}) = 0; \\
D_{[i} D_{i_{1}} \Psi_{i_{2}...i_{d-3}]} = 0, \nonumber 
\end{eqnarray} 
or in form notation as
\begin{equation}
G = f \bar{G}, \bar{G} = d\Psi \hspace{3mm} \rightarrow  
\hspace{3mm} d(* f d\Psi) = 0, dd\Psi = 0,
\end{equation}
where we have introduced the $(d-3)$ form $\bar{G}$ which is
related to $G$ by the function 
\begin{equation}
f =  \exp(\frac{4\phi\sqrt{d-2}}{(d-3)}). 
\end{equation}
We have so far simply followed the prescription of \cite{GH}, but we
find here an important difference. Although
the local existence of the potential $\Psi$ is ensured by the
closure of the $(d-3)$-form $\bar{G}$, if the periods of $\bar{G}$ 
are non-zero, the potential $\Psi$ cannot be defined globally. Even
if the $(d-3)$ cohomology of $M$ is trivial we cannot guarantee that
$\bar{G}$ has zero periods, since it is defined within the orbit space
$\Sigma$. Since in four dimensions $\Psi$ is a 
scalar, this problem did not arise in the discussions of \cite{GH}. 

\bigskip

Suppose $\bar{G}$ has non-trivial periods; we can partition the $(d-1)$
dimensional manifold $\Sigma$ into a finite
set of neighbourhoods $\sigma_{m}$ with $(d-2)$ dimensional 
boundaries $\partial \sigma_{m}$, such that each point in $\Sigma$ is
covered by a finite number of $\sigma_{m}$. Although the original $d$
dimensional manifold has no boundary by definition the $(d-1)$
dimensional manifold $\Sigma$ will
have boundaries at the fixed points of the circle action; the total
boundary $\partial \Sigma$
consists of a disjoint set of boundaries around each fixed point set. 
Contributions to the boundaries $\partial \sigma_{m}$ thus
arise both from the boundary of $\Sigma$ and from the boundaries dividing
the $\sigma_{m}$. 

Within each of the
$\sigma_{m}$ we may define a $(d-4)$ potential $\Psi_{m}$ such that in the 
overlap $\cap_{\sigma_{m,n}}$ 
between two neighbourhoods $\sigma_{m}$ and $\sigma_{n}$ the
potentials are related by gauge transformations
\begin{equation}
\Psi_{m} - \Psi_{n} = d \omega_{mn},
\end{equation}
where $\omega_{mn}$ is a $(d-5)$ form. Then we can for example express
an integral over the entire $(d-1)$ dimensional manifold in terms of
integrals over the boundaries of each neighbourhood
\begin{equation}
\int_{\Sigma} F \wedge \bar{G} = 
\sum_{m} \int_{\sigma_{m}} F \wedge d\Psi_{m}
= \sum_{m} \int_{\partial \sigma_{m}} F \wedge \Psi_{m}. \label{hmm}
\end{equation}
We will find that this particular integral arises below. 
For example, in the simplest non-trivial case, where we 
divide $\Sigma$ into two regions, each of which contains a 
single fixed point set, we find that
\begin{equation}
\int_{\Sigma} F \wedge \bar{G} =  \int_{M_{1}^{d-2}} F \wedge \Psi_{1}
+  \int_{M_{2}^{d-2}} F \wedge \Psi_{2} 
+ \int_{\partial\sigma_{1} \subset \cap_{\sigma_{1,2}}} 
F \wedge (\Psi_{1} - \Psi_{2}),
\end{equation}
where in the last term we have taken account of the opposite
orientations of the 
boundaries, and the $M_{i}^{d-2}$ are arbitrary surfaces enclosing the
fixed point sets. Thus the total integral can be related to integrals 
over the two fixed
point sets, and to a Dirac string type contribution. Note that
although the original integral is manifestly independent of the gauge
choices for the potentials, individual contributions to the integral
will depend on each gauge choice. 

\bigskip

Following the approach of \cite{GH}, we next
look for symmetries of the terms in the
Lagrangian depending on the potential and the dilaton under 
global transformations. Although $\Psi$ cannot necessarily be defined 
globally, we will find that looking for symmetries of the action
will help to indicate a well-defined way to usefully rewrite the action. 
There is a manifest symmetry under translations of the form
\begin{equation}
\bar{G} \rightarrow \bar{G} + d{\cal{A}},
\end{equation}
where $\cal{A}$ is an arbitrary exact $(d-4)$ form, and
all other fields are held constant; the associated
$(d-4)$ form translational Noether current is 
\begin{equation}
J_{T} = \bar{G}.
\end{equation}
This is simply a statement that the potentials are only defined modulo
exact forms, as we discussed above.

There is also a symmetry under a global dilation of the form
\begin{equation}
\Psi_{m} \rightarrow 
b \Psi_{m}, \hspace{3mm}
e^{\frac{4\sqrt{d-2}}{d-3}\phi} \rightarrow b^{-2} 
e^{\frac{4\sqrt{d-2}}{d-3}\phi},
\end{equation}
with the associated Noether current within each region being
\begin{equation}
J_{m(D)} = \frac{2}{\sqrt{d-2}} (\ast d\phi) + \frac{1}{2} F \wedge
\Psi_{m}.
\end{equation}
Again, if the periods of $\bar{G}$ are non-zero the dilation current is
defined locally within each submanifold $\sigma_{m}$; the
dilation currents in the intersections of different regions are
related as
\begin{equation}
J_{m(D)} - J_{n(D)} = \frac{1}{2} F \wedge d\omega_{mn}.
\end{equation} 
In the absence of the cosmological term, both dilations and translations 
are symmetries of the effective action, and the Noether currents are 
conserved, but in the presence of a cosmological term 
the symmetry under the dilation current is broken. That is, using the
field equations, we find that
\begin{eqnarray}
H_{D} &=& dJ_{D} = \Lambda e^{\frac{4\phi}{\sqrt{d-2}(d-3)}} \eta_{d-1}, \\
&=&  \frac{2}{\sqrt{d-2}} d(\ast d\phi) 
+ \frac{1}{2} \int_{\Sigma} F \wedge \bar{G}, \nonumber 
\end{eqnarray}
where $\eta_{d-1}$ is the volume form on $\Sigma$. Note that $H_{D}$
is totally independent of potentials and is well defined throughout
the $(d-1)$-dimensional manifold. In fact the existence of $H_{D}$ is
implied by the field equation for the dilaton derived from the
original action (\ref{orac}), as is easily seen if we rewrite
(\ref{orac}) in form notation. For comparison with the work of
\cite{GH}, we have derived the existence of $H_{D}$ by introducing a
dilation of the reduced action, but $H_{D}$ is defined even when
$\bar{G}$ has non-zero periods.  

\bigskip

This relationship between the cosmological constant and $H_{D}$ 
can then be used to rewrite the on-shell action for a compact
manifold without boundary as
\begin{eqnarray}
S_{E} &=& - \frac{1}{8 \pi G_{d}} \int_{M} \eta_{d} \Lambda;
\nonumber \\
 &=& - \frac{\beta}{8 \pi G_{d}} \int_{\Sigma} \eta_{d-1} (\Lambda
e^{\frac{4\phi}{\sqrt{d-2}(d-3)}}); \label{a1} \\
 &=& - \frac{\beta}{8 \pi G_{d}} \int_{\Sigma} H_{D}, \nonumber 
\end{eqnarray}
where in the first equality we express the volume form of $M$ as
$\eta_{d}$. So using the explicit form for $H_{D}$ we find that
\begin{equation}
S_{E} = - \frac{\beta}{8 \pi G_{d}} \lbrace \frac{2}{\sqrt{d-2}} \int_{\Sigma}
d(\ast d\phi) + \frac{1}{2} \int_{\Sigma} F \wedge \bar{G} \rbrace. 
\label{redac}
\end{equation}
We defer discussion of the second term to the following section; 
the first term can be related to the $(d-2)$ volumes of the fixed point
sets as follows. Since this term is globally exact it
can be converted into an integral over the boundaries
of the manifold $\Sigma$, that is, to an integral over the fixed point
set boundaries. Thus we may introduce invariant quantities, the dilation
charges, such that at the $a$th fixed point set 
\begin{equation}
M_{a} = \frac{2}{\sqrt{d-2}} \int_{M_{a}^{d-2}} d^{d-2}x \sqrt{c} 
(n \cdot \partial \phi), 
\end{equation}
where we integrate over any $(d-2)$ dimensional 
boundary surrounding the fixed point set. All physical quantities
are of course independent of the particular choice of $(d-2)$ 
dimensional manifold around each fixed point set; all surfaces
surrounding the nut or bolt that can be continuously transformed into
one another will give the same action. 

It is useful at this point to rewrite the dilation charges in terms of a
conformally rescaled metric; that is, we decompose the $d$ dimensional
metric as 
\begin{equation}
ds^2 = e^{-\frac{4\phi}{\sqrt{d-2}}}(dx^d + A_{i}dx^{i})^2  +
\tilde{g}_{ij} dx^{i}dx^{j}, \label{crm}
\end{equation}
where $\tilde{g}$ is conformally related to the metric $g$ given in 
(\ref{line-el}). $M_{a}$ can then be expressed as 
\begin{equation}
M_{a} =  \int_{M_{a}^{d-2}} d^{d-2}x \sqrt{\tilde{c}} (\tilde{n}
\cdot {\sqrt{\hat{g}_{dd}}}), \label{maz}
\end{equation}
where $\tilde{c}$ is the induced (conformally rescaled) metric on the 
boundary and $\tilde{n}$ is the normal to the boundary. 

Expressed in this form, it is evident that this term vanishes 
except when the fixed point set is of dimension $(d-2)$. The integral
must be independent of the choice of boundary around the fixed point
set. So we can take an arbitrary boundary and then take the limit that 
it is the boundary of the fixed point set itself, which necessarily 
has a vanishing $(d-2)$ dimensional volume element. As we take this
limit, the normal derivative is finite, since by definition
$\hat{g}_{dd}$ vanishes on the fixed point set, but is
non-zero on any boundary surrounding the fixed point set. Hence the
integral must vanish, unless the $(d-2)$ dimensional volume of the fixed
point set is non-zero. 

It is straightforward to evaluate the term for a $(d-2)$ dimensional
fixed point set. In the neighbourhood of the bolt, we can express the
metric in the form
\begin{equation}
ds^2 = d\rho^2 + \rho^2(d(\frac{x^{d}}{\mu}) + 
A_{\alpha}dx^{\alpha})^2 + g_{\alpha\beta}dx^{\alpha}dx^{\beta}, \label{d-2}
\end{equation}
where the periodicity of $x^d$ is as usual $2\pi\mu$. 
The $(d-2)$ dimensional metric $g_{\alpha\beta}$ is independent of
$\rho$, and $A$ is pure
gauge if the normal bundle over the bolt is trivial. The boundary of
the fixed point set is at the origin $\rho = 0$, and we choose the
boundary in (\ref{maz}) to be the fixed point set itself. Then 
\begin{equation}
M_{a} = \frac{1}{\mu} \int_{M_{a}^{d-2}} d^{d-2}x \sqrt{g} =
\frac{V_{a}}{\mu}, \label{potgd}
\end{equation}
where $V_{a}$ is the $(d-2)$ volume of the fixed point set, evaluated in
the original metric. This gives the intrinsic contribution to the
action from $(d-2)$ dimensional bolts. So the total action becomes:
\begin{equation}
S_{E} = - \sum_{a} \frac{V_{a}}{4 G_{d}} - \frac{\beta}{16 \pi G_{d}}
\int_{\Sigma} F \wedge \bar{G},
\end{equation} 
where only $(d-2)$ dimensional fixed point sets contribute to the
first term. We hence find that there are intrinsic contributions to
the action from the $(d-2)$ volumes of the fixed point sets. Another
way of stating this is to say that there is a contribution to the
action from the volume of the boundary of the orbit space. There are
additional contributions arising from the nut behaviour of the
fixed point sets and from non-trivial $(d-3)$ cohomology of $\Sigma$
which we shall now discuss.

\section{Cohomology contributions to the action} \label{pca}
\noindent

In four dimensions, our expression for the action reduces to (\ref{4dent})
in agreement with \cite{GH}. Since we can write $\bar{G} = d \Psi$
globally, where $\Psi$ is a scalar function, we can express the
integral over $\Sigma$ as an integral over only boundaries of fixed point
sets and thus 
\begin{equation}
S_{E} = - \sum_{a} \frac{V_{a}}{4 G_{4}} - \frac{\beta}{16 \pi G_{4}}
\sum_{a} \int_{M_{a}^{2}} F \wedge \Psi.
\end{equation} 
The potential is a scalar function and so 
\begin{equation}
\int_{M_{a}^{2}} F \wedge \Psi = \Psi_{a} \int_{M_{a}^{2}} F,
\end{equation}
i.e. the integral over the potential terms in the action reduces to an
integral of the 2-form over a surface surrounding the fixed point
set. This integral is related to the first Chern number of the $U(1)$
bundle over the space of non-trivial orbits, and, as is discussed in
\cite{GH}, one can show that the nut charge is given by $\beta/8\pi pq$
for a nut of type $(p,q)$, and by ${\cal{Y}}/8\pi$ for a bolt of
self-intersection number $\cal{Y}$. 

\bigskip

In higher dimensions we cannot in general reduce the integral over
$\Sigma$ to integrals over only fixed point sets; there can also be
contributions related to the non-trivial $(d-3)$ cohomology of $\Sigma$. 
We will postpone the discussion of the general case, and assume
that $\bar{G}$ has zero periods so that we can introduce a global potential 
$\Psi$. Then the integral over $\Sigma$ becomes
\begin{equation}
- \frac{\beta}{16 \pi G_{d}} \int_{\Sigma} F \wedge \bar{G} = 
- \sum_{a} \frac{\beta}{16 \pi G_{d}} \int_{M_{a}^{d-2}} F \wedge \Psi,
\label{pft}
\end{equation}
where we take the integrals over $(d-2)$ dimensional manifolds
surrounding each fixed point set. Thus we can associate contributions
to the action from the nut behaviour of each fixed point set. As in
four dimensions, the total contribution is gauge invariant, but
individual contributions do depend on the choice of gauge.

The form of (\ref{pft}) will be
particularly simple when the gauge field is independent of the
coordinates of the fixed point set. That is, if the fixed point set
can be surrounded by a $(d-2)$ manifold which is the product of the 
the $(d-2k)$ dimensional fixed point set, and a $(2k-2)$ dimensional 
hypersurface of small characteristic size $\epsilon$, the form of the
integral simplifies because the metric is a product metric. In
physical terms, the requirement is that there are no Dirac string type
singularities associated with the fixed point set in the $d$
dimensional manifold. So such a decomposition will always be possible
when the second cohomology class of the $d$ dimensional manifold is
trivial. 

We can show this as follows; if in the neighbourhood of the fixed
point set the metric can be expressed as a product of a $(d-2k)$
dimensional metric $g(x^{i})$ and a $(2k-2)$ dimensional 
metric $g(\theta^{i})$, then $\bar{G}$ is also a product
\begin{equation}
\bar{G} = \bar{G}_{1}(\theta^{i}) \wedge \bar{G}_{2}(x^{i}).
\end{equation}
That is, the dual
field strength can be expressed as the exterior product of a $(d-2k)$
form and a $(2k-3)$ form, whose only non vanishing components are the 
$\theta^{i}$ and $x^{i}$ components respectively. The potential can also
be expressed as the exterior derivative 
$\Psi = \Psi_{1}(\theta^{i}) \wedge \Psi_{2}(x^{i})$ where $\Psi_{1}$
is a $(2k-4)$ form and $\Psi_{2}$ is a $(d-2k)$ form. Then,
\begin{equation}
d\Psi_{1} \wedge \Psi_{2} + \Psi_{1} \wedge d\Psi_{2} = \bar{G}_{1}
\wedge \bar{G}_{2}.
\end{equation}
Now it is easy to see that $\bar{G}_{2} = \Psi_{2}$ is closed 
and $\bar{G}_{1} = d\Psi_{1}$ and so 
\begin{equation}
\int_{M^{d-2}} F \wedge \Psi = \int_{M^{2k-2}}F \wedge \Psi_{1}
\int_{M^{d-2k}} \bar{G}_{2}.
\end{equation}
Furthermore, $\bar{G}_{2}$ is the volume form of the fixed point set,
and hence,
\begin{equation}
\int_{M^{d-2}} F \wedge \Psi = V_{d-2k} \int_{M^{2k-2}} F \wedge \Psi_{1}, 
\end{equation}
where $V_{d-2k}$ is the $(d-2k)$ volume of the fixed point set in the
metric $g(x^{i})$. 

For a $(d-4)$ dimensional fixed point set, the integral reduces 
further to 
\begin{equation}
\int_{M^{d-2}} F \wedge \Psi = \Psi_{f} V_{d-4} \int_{M^2} F,
\end{equation}
where $\Psi_{f}$ is a scalar function, evaluated at the fixed point set. 
The integral is as usual given by $\beta/pq$ where the normal
bundle over the fixed point set is characterised by the two integers 
($p,q$), and hence
\begin{equation}
- \frac{\beta}{16 \pi G_{d}} \int_{M^{d-2}} F \wedge \Psi 
= -\frac{\beta^2}{16 \pi pq G_{d}} \Psi_{f} V_{d-4}, \label{esa}
\end{equation}
which gives the required answer in four dimensions. 

\bigskip

We can also obtain this answer by working with an explicit form of the metric.
In a small
neighbourhood of the fixed point set, we can always express the metric
in the form
\begin{equation}
ds^2 = (d\rho_{1}^2 + \rho_{1}^2 d\psi_{1}^2) + (d\rho_{2}^{2} +
\rho_{2}^{2}d\psi_{2}^2) + d\bar{s}_{d-4}^2,
\end{equation}
since the normal bundle can always be locally be decomposed into a sum
of complex line bundles. If the second cohomology class is trivial
such a decomposition is valid throughout the neighbourhood of the
fixed point set. In general, although locally we can bring the metric
into this form, there will be non-trivial mappings between different
neighbourhoods of the fixed point set. This will always be so if the
second cohomology of the original $d$-dimensional manifold is
non-trivial. 

We take the Killing vector
to be $k = \partial_{\psi_{1}} + \partial_{\psi_{2}}$ which has a zero at 
$\rho_{1} = \rho_{2} = 0$ and introduce the coordinate
$\bar{\psi}_{2} = \psi_{2} - \psi_{1}$ such that
$\partial_{\bar{\psi}_{2}}$ is invariant along orbits of
$\partial_{\psi_{1}}$. We then find that
\begin{equation}
ds^2 = (d\rho_{1}^2 + \rho_{1}^2 d\psi_{1}^2) + (d\rho_{2}^{2} +
\rho_{2}^{2}(d\bar{\psi}_{2} + d\psi_{1})^2) + d\bar{s}_{d-4}^2.
\end{equation}
We introduce new coordinates
\begin{equation}
\rho_{1} = \epsilon \cos\theta, \hspace{5mm} \rho_{2} =
\epsilon\sin\theta,
\end{equation}
where the range of the angular coordinate is from $0$ to $\pi/2$. 
Dimensionally reducing, we find that the $(d-1)$ dimensional fields are 
\begin{eqnarray}
ds^{2} &=& d\epsilon^2 + \epsilon^2d\theta^2 + \epsilon^2 \sin^2\theta
\cos^2 \theta d\bar{\psi}_{2}^2 + d\bar{s}_{d-4}^2; \nonumber \\
A_{\bar{\psi}_{2}} &=& \sin^2\theta; \\
g_{\psi_{1}\psi_{1}} &=& \epsilon^2 = \exp(-{4\phi}/{\sqrt{d-2}}),
\nonumber 
\end{eqnarray}
where we give the conformally rescaled metric defined in (\ref{crm})
for notational simplicity. We can then show that
\begin{equation}
F_{\theta\bar{\psi}_{2}} = 2\sin\theta\cos\theta,
\end{equation}
is the only independent component of the gauge field strength. 
Dualising the two form field, we find that the only independent
component of the $(d-3)$ form field is
\begin{equation}
G_{i_{1}...i_{d-4}\epsilon} = \epsilon^{\frac{1-d}{d-3}}
\sqrt{g_{d-4}},
\end{equation}
where $g_{d-4}$ is the determinant of the metric on the fixed point
set in the conformally rescaled metric. Using the relation between
this field and the potential, we can extract the form of the potential
in the vicinity of the fixed point sets as
\begin{equation}
\Psi = (\Psi_{f} + O(\epsilon^2))\eta,
\end{equation}
where $\Psi_{f}$ is a constant, and $\eta$ is the volume form of the
fixed point set. 
Now the potential term in the action can be written as in (\ref{pft})
where $M^{d-2}$ is any $(d-2)$ manifold surrounding the fixed point
set; thus choosing it to be the product manifold of the $(d-4)$
dimensional fixed point set and a surface of constant $\epsilon
\rightarrow 0$, we find
\begin{equation}
S_{E}^{(2)} = - \frac{\beta^2}{16 \pi G_{d}} \Psi_{f} V_{d-4}.
\end{equation}
where in the limit $\epsilon \rightarrow 0$ this is the only
contributing term.
So we have explicitly shown that the integral may be decomposed as 
in (\ref{esa}), where we set $p,q=1$; it is straightforward to extend
the proof to general integers by taking a Killing vector $k = p
\partial_{\psi_{1}} + q \partial_{\psi_{2}}$. 

\bigskip

Returning to the general case, when $\bar{G}$ has non-zero periods, 
following (\ref{hmm}) we can express the Dirac string terms in the action as 
\begin{equation}
- \frac{\beta}{16 \pi G_{d}} \int_{\Sigma} F \wedge \bar{G} = 
- \frac{\beta}{16 \pi G_{d}} \lbrace \sum_{a} \int_{M_{a}^{d-2}} F
\wedge \Psi_{a} + \sum_{m<n} \int_{\partial\sigma_{m} \subset 
\cap \sigma_{m,n}} F \wedge d\omega_{mn} \rbrace. 
\end{equation}
That is, we decompose the integral over the entire manifold into
contributions from each fixed point set, and from the Dirac string
type behaviour of the $(d-1)$-dimensional manifold. We have implicitly
assumed here that we can introduce a single potential
within the neighbourhood of each fixed point set but it is
straightforward to relax this condition.
Note that even if one can further decompose the integral of $F \wedge
\Psi$ at individual fixed point sets, one still has to allow for 
the non-trivial $(d-3)$ cohomology of the $(d-1)$-dimensional
manifold, and the integrals do not take simple forms. 

Since under gauge transformations of the potentials both the fixed
point set and Dirac string terms change it is more useful to evaluate
the total contribution to the action from the nut behaviour and
$(d-3)$ cohomology; one cannot associate a gauge invariant
contribution to the action from the nut behaviour of any one fixed
point set. 

In decomposing the integral over $\Sigma$ we have used the fact that 
$F \wedge \bar{G}$ is closed to introduce a local potential $F \wedge
\Psi$. We could of course introduce a local potential $A \wedge \bar{G}$
instead. By definition, $F$ is not globally exact; if it were we could
gauge transform our original circle coordinate $\tau$ and remove all
gauge field contributions. So on introducing a local potential of this
form we would still have to partition the orbit space
and define the transformations of the
potentials between regions. This illustrates further that there is
little meaning in identifying contributions to the action
from particular parts of $\Sigma$ particularly when $\bar{G}$ is not globally
exact. 

\bigskip

It is interesting to consider a class of compact Einstein manifolds
admitting no non trivial fixed point sets. One can 
regard the $(2n+1)$ sphere as a $U(1)$ bundle over $CP^n$ with
the action of the $U(1)$ being trivial in the sense of having no fixed
point sets. The Kaluza Klein two form is the unique self dual two form in
$CP^n$ and the action of the dilaton is trivial. The dual field is
defined by $G =
\ast F = F^{n-1}$, which necessarily has non trivial periods, and thence the
potential $\Psi$ is not well defined globally. 

Since there are no fixed point sets, the boundary of $\Sigma$, the set
of boundaries of the fixed point sets, vanishes. We can integrate the
potential term over $\Sigma$ to find 
\begin{equation}
S_{E} = - \frac{\beta}{8 \pi G_{d}} \int_{\Sigma} (F \wedge \ast F),
\end{equation}
and the integral reduces to the volume of the base
manifold. Evaluation of this integral by the division of $\Sigma$ into
regions and the introduction of potentials would give the same answer.
Now this compares to an action which we can explicitly evaluate to be
\begin{equation}
S_{E} = - \frac{\Lambda}{8\pi G_{d}} V_{2n+1}(S^{2n+1}).
\end{equation}
These two expressions appear different. However, if we chose the
standard $SU(n+1)$ invariant metric on the sphere, the metric will not
be Einstein. We thus choose on the sphere the canonical metric with
a constant curvature of one, which {\it is} $SU(n+1)$ invariant, so that the
fibration has totally geodesic fibres onto the symmetric metric on the
base manifold \cite{Be}. The fibre is a great circle of the sphere, with
length $2 \pi$, and one can then see that the two expressions
for the action are equivalent.

\bigskip

Let us consider two further examples for which $\bar{G}$ is globally exact. 
We discuss first
the $d$-dimensional sphere, endowed with canonical metric, of radius
$(d-1)^{1/2}\Lambda^{1/2}$, which satisfies $R_{ij} = \Lambda g_{ij}$.
Evaluating the action directly from (\ref{a1}) we find that 
\begin{equation}
S_{E} = - \frac{\Lambda}{8 \pi G_{d}} V_{d}(\sqrt{\frac{d-1}{\Lambda}}),
\end{equation}
where $V_{d}(a) =  2 \pi^{(d+1)/2}
a^d/\Gamma[\frac{1}{2}(d+1)]$ is the volume of a 
$d$-dimensional sphere of radius $a$. 

The action of a rank 2 generator of the $SO(d+1)$
isometry group will leave fixed a single $(d-2)$ dimensional spherical
bolt. There is no contribution to the action from the potential term,
since the second cohomology class is trivial, and hence the 
field in (\ref{d-2}) is pure gauge. So we can obtain the action as:
\begin{equation}
S_{E} = - \frac{V_{(d-2)}(\sqrt{\frac{d-1}{\Lambda}})}{4 G_{d}},
\end{equation}
which is equivalent to the previous expression. 

The fixed points of the 
action of a rank 4 or higher generator of the $SO(d+1)$ isometry group
can be regarded as the intersection of
the $(d-2)$ dimensional fixed point sets of independent generators of
the Lie algebra; that is, we can decompose the circle action as $q =
\sum_{i} n_{i} \partial_{\psi_{i}}$. 
The action evaluated using the fixed points of such a circle subgroup
will be a potential term specified entirely by 
the periodicity of the action, these integers 
and the volume of the fixed point set. The Lefschetz fixed point
theorem tells us that the action of a rank $2k$ generator leaves
fixed a single sphere of dimension $(d-2k)$, or two points if $d=2k$.

\bigskip

If we consider radially extended $U(1)$ bundles over compact
homogeneous manifolds, such as those constructed in \cite{PP},
we can progress further in the evaluation of
the potential term. The simplest example is a 
complex projective space $CP^n$ of real
dimension $d=2n$; although such instantons seem to have little
physical relevance, since there exists no Lorentzian continuation,
they illustrate several important points. Suppose that
\begin{equation}
g = \lbrace e^{i\xi_{0}},...e^{i\xi_{n}} \rbrace,
\end{equation}
is an element of the torus group $T^{n+1}$ acting on the complex
coordinates $z_{i}$ of $CP^n$ as:
\begin{equation}
(e^{i\xi_{0}},..,e^{i\xi_{n}}) \cdot (z_{0},..,z_{n}) = 
(e^{i\xi_{0}} z_{0},..,e^{i\xi_{n}} z_{n}),
\end{equation}
where the definition of complex projective space is that
$(z_{0},..,z_{n}) \in \lbrace C^{n+1} - \lbrace 0 \rbrace \rbrace /C = 
CP^{n}$. The action of $g$ leaves a point fixed if 
\begin{equation}
(e^{i\xi_{0}} z_{0},..,e^{i\xi_{n}} z_{n}) = (z_{0},..,z_{n}).
\end{equation}
This requires that 
\begin{equation}
e^{i \xi_{k}} z_{k} = e^{i \xi} z_{k} \hspace{.5in} k = 0,..,s
\end{equation}
for some $\xi$ which is determined uniquely since at least
one of the $z_{k}$ is non-zero. In fact, $\xi$ must equal one of the $\xi_{k}$.
Then we can express the manifold $X^{g}$ which is fixed under the
action of $g$ as $X^{g} = \cup X(\xi)$, with 
\begin{equation}
X(\xi) = \lbrace (z_{0},..,z_{n}) \in CP^n : 
\xi_{k} \neq \xi \rightarrow z_{k} = 0, \rbrace
\end{equation}
Thus the action of the isometry leaves fixed a set of complex 
projective spaces of various dimensions.

Now, for a complex projective space $CP^n$, the odd Betti numbers vanish
and the even Betti numbers are all equal to one; so the Euler
characteristic is given by $(n+1)$. From the Lefschetz fixed point
theorem (\ref{eul}), we can then constrain the fixed point sets of the group
action. For example, if an element of the isometry group leaves
invariant a submanifold isomorphic to $CP^{n-1}$, it must also leave 
invariant a single point. The natural interpretation of the action of
this element is that it leaves fixed the origin 
$(z_{0},...,0)$ and the $CP^{n-1}$ submanifold at
``infinity'' $(0,z_{1},..,z_{n})$, where we use quotation marks
because the manifold is of course compact.
Evidently we may similarly constrain 
the action of other elements of the isometry group. 
In particular, there must exist a generator $g$ contained in the
isometry group which has $(n+1)$ nut fixed point sets, at the
origin $(z_{0}:...:0)$ and at the ``poles at infinity'' 
$(0:..:z_{i}:..:0)$.

\bigskip

We can express the metric on $CP^n$ in the following way.
Constructing a $U(1)$ bundle over $CP^{n-1}$, with its standard Fubini
Study Einstein-K\"{a}hler metric, we obtain \cite{PP}
\begin{equation}
ds^{2} = 2(n+1) \Lambda^{-1} \lbrace d\theta^{2} + \sin^{2}\theta 
\cos^{2}\theta (d\tau - A)^{2} + \sin^{2}\theta d\bar{s}_{2(n-1)}^{2}
\rbrace, \label{clmet}
\end{equation}
with endpoints at $\theta = 0$ and $\theta = \pi/2$. We choose the
normalisation of the metric on $CP^{n-1}$ such that $R_{ij} = 2n
g_{ij}$, and $dA$ can be chosen as the K\"{a}hler form on $CP^{n-1}$. 
Then the resulting metric is isometric to the standard Fubini Study metric on 
$CP^{n}$. In particular, the Killing vector $\partial_{\tau}$ has a 
nut at the ``origin'' $\theta = 0$ and a $CP^{n-1}$ bolt at ``infinity''
$\theta = \pi/2$. The period of this circle action is $2 \pi$.  

This form of the metric is particularly useful in the evaluation of
the action; we find that
\begin{equation}
S_{E} = - \frac{\Lambda}{8 \pi G_{d}} V_{d}(M),
\end{equation}
where $V_{d}(M)$ denotes the volume of the $CP^n$ in the metric 
(\ref{clmet}), or explicitly, 
\begin{equation}
S_{E} = - \frac{d}{8n G_{d}} (\frac{d}{\Lambda})^{n-1}
V_{2(n-1)}, \label{smx}
\end{equation}
where $V_{2(n-1)}$ is the volume of the base manifold. It can be
verified that the bolt at
infinity contributes a volume term to the action
\begin{equation}
S_{E}^{(1)} = - \frac{1}{4 G_{d}} (\frac{d}{\Lambda})^{n-1}
V_{2(n-1)},
\end{equation}
whilst the cohomology of $\Sigma$ contributes a term 
\begin{equation}
S_{E}^{(2)} = - \frac{1}{4n G_{d}} (\frac{d}{\Lambda})^{n-1}
V_{2(n-1)}, \label{xzy}
\end{equation}
and thence the two contributions do sum to (\ref{smx}) as
required. This provides a verification that our decomposition of the action  
holds; the reason for choosing projective spaces as an example derives
from the simplification in the integral over $\Sigma$ because of 
the form of the cohomology structure for such spaces. 
Part of the integral reduces to
an integral of $F \wedge *F$ over $CP^{n-1}$, where we take the dual
in the metric on the base space. Since there is only one
independent closed but non exact (${n-1}$) form, the integral is then
proportional to the volume of the base space. 

Since the $(d-3)$ form
$\bar{G}$ has trivial cohomology, one can find a potential
globally of the form 
\begin{equation}
\Psi = \gamma(\theta,\Lambda) (*F),
\end{equation}
where we take the dual in the metric on $CP^{n-1}$; 
the function $\gamma$ can of course be explicitly determined. One can
add an arbitrary constant to this function to ensure that it vanishes
either at the origin or at ``infinity'', with the latter being the more
natural choice. Thus we can convert the volume integral over $\Sigma$
to integrals over the boundaries of the fixed point sets; for
instance, in the vicinity of the bolt, the form of the reduced metric is
\begin{equation}
ds^2 \propto (d\epsilon^2 + d\bar{s}_{2n}^2),
\end{equation}
and we can take the surrounding surface
$M^{2n}$ to be surface of arbitrary small $\epsilon$. The 
two integrals over the fixed point sets will sum to (\ref{xzy}) but
the individual contribution from each will depend on the gauge choice for 
$\Psi$.

\section{Entropy and the cosmological interpretation} \label{eci}
\noindent

For vacuum gravity, our results have an interpretation in 
terms of the entropy associated with the fixed point sets and Dirac
string behaviour. The derivation of the gravitational
entropy of compact solutions is perhaps less familiar than that
of black hole solutions; following \cite{GH} we introduce a 
partition function $Z$ for the canonical ensemble
\begin{equation}
Z = \sum_{m} \left \langle g_{m}|g_{m} \right \rangle,
\end{equation}
where $|g_{m} \rangle $ is an orthonormal basis of states 
for the gravitational field with a given value of $\Lambda$. 
It is important to realise that for compact solutions
there is no externally imposed temperature or potential; the
probability of each state is then $p_{m} = Z^{-1}$ and the entropy is
given by
\begin{equation}
S = \sum_{m} p_{m} \ln p_{m} = \ln Z.
\end{equation}
As usual, by the stationary phase approximation, one would expect
the dominant contribution to the partition function $Z$ which is
defined as:
\begin{equation}
Z = \int d[\hat{g}] \ e^{-S_{E}[\hat{g}]},
\end{equation}
to come from metrics near a solution $\hat{g}_{0}$ of the classical 
field equations and thus the value of $\ln Z$ to be approximately
$S_{E}[\hat{g}_{0}]$. Then the entropy will be
given by:
\begin{equation}
S =  \sum_{a} \frac{V_{a}}{4 G_{d}} + \frac{\beta}{16 \pi G_{d}}
\int_{\Sigma} F \wedge \bar{G}. 
\end{equation} 
Hence not only do the $(d-2)$ bolts have an entropy equal to a quarter
of their $(d-2)$ volume, but there is also a contribution to the
entropy from the nut behaviour of the fixed point sets, and from the
$(d-3)$ cohomology of $\Sigma$. This gives the extension to general 
dimensions of the result given in \cite{GH}. 

\bigskip

The cosmological relevance of this result is as follows; according to
the no boundary proposal, the quantum state of the universe is defined
by path integrals over Euclidean metrics on compact manifolds $M$. One
usually considers this proposal in a four dimensional context, but the
same ideas follow for higher dimensional theories. We give only a
brief summary of the procedure here; further discussion may be found
in \cite{EQG}. The subsequent
Lorentzian evolution is described by initial data on a zero momentum
hypersurface $\Sigma_{i}$ of dimension $(d-1)$. If $M$ is simply
connected, then the hypersurface divides the manifold into two parts,
$M_{\pm}$, which are usually assumed to have equal action. 

One then defines a path integral over all metrics on $M_{+}$ that
agree with the induced metric on the hypersurface $\Sigma_{i}$; this gives
the wavefunction of the universe $\Psi(b_{ij})$
\begin{equation}
\Psi(b_{ij}) = \int d[g] \exp(-S_{E}(g)),
\end{equation}
where $b$ is the induced metric on the boundary. The Euclidean action
is given by 
\begin{equation}
S_{E} = -\frac{1}{16\pi G_{d}} \int_{M_{+}} d^dx \sqrt{\hat{g}} [R - m]
- \frac{1}{8\pi G_{d}} \int_{\Sigma_{i}} d^{d-1}x \sqrt{b} 
{\cal{K}}_{\Sigma_{i}},
\end{equation}
where ${\cal K}_{\Sigma_{i}}$ is the trace of the second 
fundamental form on the boundary $\Sigma_{i}$; since we regard 
the hypersurface as initial
data for the subsequent Lorentzian evolution, the hypersurface must have
zero momentum, and this geometry term vanishes. 

The absence of any externally imposed temperature
allows us to define a partition function, $Z = \left | \Psi(b_{ij}) \right
|^2$, which we interpret as a probability of the process occurring, 
and an entropy which is given by
\begin{equation}
S = -2 S_{E}.
\end{equation}
Now one can decompose the volume term in the original action in terms of the
fixed point sets of a Killing vector; we are assuming that the manifold
can be divided symmetrically, so that the action of $M_{+}$ is equal
to half the action of $M$. Thence the entropy will be given by the sum
over the fixed point sets in the original compact manifold; in the
subsequent Lorentzian evolution, these 
fixed point sets will have the interpretation of, for example, black
hole horizons. 

\bigskip

We give two examples here; creation of a $d$ dimensional universe with
positive cosmological constant is described by a $d$ sphere. For
definiteness, we consider the five dimensional sphere with its standard 
round metric
\begin{equation}
ds_{5}^{2} = d\chi^2 + \cos^2\chi [ d\rho^2 + \frac{1}{4}\sin^2\rho
((d\theta^2 + \sin^2\theta d\psi^2) + (dx^{5} + \cos\theta d\psi)^2) ],
\end{equation}
where $-\pi/2 \le \chi \le \pi/2$, $0 \le \rho \le \pi$, $0 \le \theta
\le \pi$, $0 \le \psi \le 2\pi$ and $0 \le x^{5} \le 4\pi$. Then we can
calculate the action directly to be
\begin{equation}
S_{E} = - \frac{\pi^2}{2G_{5}}. \label{above}
\end{equation}
One can also calculate the action by looking at the fixed point sets
of the Killing vector $\partial_{x^{5}}$ which generates the Hopf
fibration on the three sphere $\rho, \chi = $ constant; there are
fixed point sets at $\rho = 0, \pi$. We can choose the
potential so that each fixed point set contributes equally to the
action, that is, so that the contribution from each is
$-\pi^2/4G_{5}$.

One can consider the creation of a five dimensional universe by taking
a tunnelling geometry $-\pi/2 \le \chi \le 0$, with the Lorentzian
section described by $\chi = i t$ with $t$ positive. This gives a five
dimensional expanding de Sitter universe. We now only include
half of each fixed point set, 
and thus the action for the
compact manifold is half of (\ref{above}), as required. The total
entropy is given by minus twice the action, and we can associate it
with contributions from each of the fixed point sets. 

If we treat the
Killing direction as compact, these fixed point sets correspond to
monopoles; the effective four-dimensional Lorentzian solution
describes pair creation of monopoles within an expanding background,
where there is an entropy associated with each monopole. This
interpretation is
discussed in \cite{CCG}; note that the example is illustrative but not
physically realistic, since the compact direction is also expanding. 

\bigskip

As a second example, we could consider the Euclidean Schwarzschild de Sitter 
solution in general dimensions, with suitable choice of parameters to 
ensure regularity. The latter choice is somewhat subtle \cite{RafH}, 
and we shall
not discuss it here; one would however expect to be able to choose an
imaginary time Killing vector with fixed point sets such that 
the action for the compact solution is
one quarter of the volumes of the black hole and cosmological
horizons.

Upon choosing an
appropriate initial value hypersurface, the subsequent Lorentzian
evolution will describe pair creation of Schwarzschild black holes
within a cosmological background. The entropy for the process will be given
by minus the original action, that is, 
one quarter of the horizon volumes; the exponential of the
entropy gives a measure of the probability of the pair creation.
We could take these ideas further; for example, 
the Page solution \cite{DPag} would be an instanton for cosmological
pair creation of Taub-Bolts.

\section{Generalisation to supergravity theories} \label{ggft}
\noindent

It is interesting to consider whether one can extend the ideas of
entropy associated with the fixed point sets of isometries to compact
solutions of theories involving not only the graviton, but also other
fields. We will consider here a generic action of the form 
\begin{equation}
S_{E} = - \frac{1}{16\pi G_{d}} \int_{M} d^d x \sqrt{\hat{g}} [R 
- e^{-b \Phi} m - (\partial\Phi)^2 - e^{-a\Phi}{\mathcal H}_{p+1}^2],
\end{equation}
where $\Phi$ is the dilaton, and ${\mathcal H}_{p+1}$ is a $(p+1)$ form. 
Depending on the values of $a$, $b$ and $p$, this will 
give the appropriate action 
for Einstein-Maxwell theories coupled to a dilaton, and for particular limits 
of supergravity theories. 
Using the field equations we can rewrite the action as
\begin{equation}
S_{E} = - \frac{1}{16\pi G_{d}} \int_{M} d^dx \sqrt{\hat{g}} 
[2 e^{-b\Phi} \Lambda -  \frac{2p}{(d-2)} e^{-a\Phi} {\cal H}_{p+1}^2].
\end{equation}
Isometries of solutions must map not just the graviton, but
also the other fields, into themselves. 
If we assume the existence of a one parameter isometry group, we can 
dimensionally reduce along closed orbits of the Killing vector and
re-express the action in terms of the $(d-1)$ dimensional fields. 

From the $d$ dimensional gauge field, we will obtain a $(d-1)$
dimensional $(p+1)$ form ${\mathcal H}_{m}$ and a $(d-1)$ dimensional 
$p$ form ${\mathcal H}_{e}$. We will call the former the ``magnetic'' part
of the field, and the latter the ``electric'' part of the field. The
reason for this terminology is that we will 
later analytically continue the solution, and interpret the Killing
direction as the imaginary time. With this interpretation, the $(d-1)$
dimensional gauge field arising from the metric must vanish if a
Lorentzian evolution is to exist, since otherwise the Lorentzian
and Euclidean metrics could not both be real. This then implies that
the imaginary time Killing vector has only $(d-2)$ dimensional fixed
point sets, which we will interpret as horizons in the Lorentzian 
continuation. 

It is perhaps unnecessary to assume that the Euclidean metric is
real. Since we allow electric gauge fields which are pure imaginary on
the Euclidean section, we should also permit the Euclidean metric to
be complex provided that the metric is real in the Lorentzian
continuation. However, few useful
complex metrics of this type are known, and we shall not consider
them here. 

\bigskip

Let us take the $(p+1)$ form to be pure magnetic. Since $F$ is pure
gauge, the potential $\Psi$ vanishes and the dilation current is
well-defined globally. We can regard  
the cosmological constant and magnetic field 
terms in the $(d-1)$ dimensional action as breaking the symmetry 
under dilations so that
\begin{equation}
d J_{D} = \eta_{d-1} \lbrace e^{-b\Phi} 
\Lambda e^{\frac{4\phi}{\sqrt{d-2}(d-3)}} - \frac{p}{(d-2)} 
e^{-a\Phi}{\cal H}^2_{p+1}  e^{-\frac{4p \phi}{\sqrt{d-2}(d-3)}}
\rbrace,
\end{equation}
where the dilation current is defined as before. Relating this to the
on shell action given above, we find that
\begin{equation}
S_{E} = - \frac{\beta}{8\pi G_{d}} \int_{\Sigma} dJ_{D}
= - \sum_{a} \frac{V_{a}}{4G_{d}},
\end{equation}
where we need only sum over $(d-2)$ dimensional fixed point sets, since we 
are assuming that the $(d-1)$ dimensional gauge field is trivial.
Note that the
gradient of the dilaton field $\Phi$ does not contribute to the 
divergence of the dilation current, and its contribution to the 
action vanishes on shell. So we can see that the entropy, which is 
still given by minus the action in the saddle-point approximation, 
since there is no externally imposed temperature, is again given 
entirely by the contributions from the fixed point sets.

If we assume instead that the field is pure electric, the action can
be decomposed in terms of the fixed point sets, and a volume integral
of the field
\begin{equation}
S_{E} =  - \sum_{a} \frac{V_{a}}{4G_{d}} + \frac{1}{8\pi G_{d}} \int_{M}
  d^dx \sqrt{\hat{g}} e^{-a\Phi} {\cal H}_{p+1}^2.
\end{equation}
That is, the action for the solution depends not only on the fixed
point sets, but also on a volume integral of the gauge field. 

\bigskip

These results again have an interesting interpretation in the
context of cosmological pair creation. Let us
take the Killing vector to be $\partial_{\tau}$, and 
interpret $\tau$ as the imaginary time; we then
divide the compact manifold along an appropriate zero momentum hypersurface.
For a pure magnetic two form, there are no
boundary contributions to the action, and the total entropy is given
by
\begin{equation}
S = \sum_{a} \frac{V_{2}}{4 G_{4}} \label{sumetr}
\end{equation}
where the summation runs over all fixed point sets of the Killing
vector $\partial_{\tau}$ in $M$ and we restrict to four dimensions
since one cannot define magnetic charge in higher dimensions.

For a pure electric field, one should however include a boundary term 
in the action to obtain the correct equations of motion when the 
electric charge is held fixed \cite{HR}. That is, we want to use an action 
whose variations give the Euclidean equations of motion when the variation 
fixes the boundary data on the initial value hypersurface $\Sigma_{i}$.
The appropriate action is then
\begin{equation}
S_{E}^{tot} = \frac{1}{2}S_{E} - \frac{1}{4 \pi G_{d}} 
\int_{\Sigma_{i}} d^{d-1}x
\sqrt{b} e^{-a\Phi} \bar{F}^{\mu\nu}n_{\mu}\bar{A}_{\nu},
\end{equation}
where $S_{E}$ is the action for the total manifold given above
and $n$ is the normal to the boundary. Note that
our notation for the gauge field is intended to differentiate between the $d$
dimensional fields, and the $(d-1)$ dimensional gauge
field we obtain upon dimensional reduction. One can  
convert the remaining Maxwell field term in the volume part of the action to 
a surface term
\begin{eqnarray}
S_{E}^{(2)} &=& \frac{1}{8 \pi G_{d}} \int_{M_{+}} d^dx \sqrt{\hat{g}} 
e^{-a\Phi} \bar{F}^{\mu\nu}D_{\mu}\bar{A}_{\nu}; \nonumber \\
&=& \frac{1}{4\pi G_{d}} \int_{\Sigma_{i}} d^{d-1}x
\sqrt{b} e^{-a\Phi} \bar{F}^{\mu\nu}n_{\mu}\bar{A}_{\nu},
\end{eqnarray}
where we have used the equation of motion for the gauge field
\begin{equation}
D_{\mu}(e^{-a\Phi} \bar{F}^{\mu\nu}) = 0.
\end{equation}
Thus, the two gauge field terms in the action cancel out, and the entropy is
again given by the summation
over the fixed point sets in $M$ (\ref{sumetr}). Such a result was
found explicitly in \cite{MaRo} for cosmological production of black
hole pairs in four dimensions; the entropy is generally given 
by the one quarter of the 
area of the black hole horizon, plus one quarter of the area of the 
cosmological horizon. Note that our analysis breaks down for
production of extreme black holes, since we would find that there was an inner 
boundary of which we would have to take account. We should also mention
that we are implicitly assuming that we can find an appropriate
non-singular choice of gauge.

\bigskip

We give as an example a particular limit of the four dimensional 
Reissner-Nordstr\"{o}m de Sitter instanton
\begin{equation}
ds^2 = \frac{1}{A} (d\chi^2 + \sin^2\chi d\psi^2) + \frac{1}{B}
(d\theta^2 + \sin^2 \theta d\phi^2),
\end{equation}
where $\chi$ and $\theta$ both run from $0$ to $\pi$, and the other
coordinates have period $2\pi$. The dilaton field is trivial; this is
a solution of Einstein-Maxwell theory. The interpretation of this
solution is of pair creation of charged black holes within a
cosmological background. This solution is obtained from
the Reissner-Nordstr\"{o}m de Sitter solution in the limit that the
black hole and cosmological horizons are at the same radius. 

The cosmological constant is given by 
$\Lambda = (A+B)/2$, and the magnetic and electric gauge fields are 
\begin{eqnarray}
\bar{F}_{magn} &=& q \sin\theta d\theta \wedge d\phi; \nonumber \\
\bar{F}_{elec} &=& -iq \frac{B}{A} \sin\chi d\chi \wedge d\psi,
\end{eqnarray}
where the magnetic/electric 
charge is defined by $q^2 = (B-A)/2 B^2$. 

One can compute the action
for the magnetic instanton by looking at the fixed point sets of 
$\partial_{\psi}$; there are two spherical fixed point sets at
$\chi = 0,\pi$, and thus the action is given by
\begin{equation}
S_{E} = - \frac{1}{4G_{4}}( 2 \times \frac{4\pi}{B} ) = -
\frac{2\pi}{B G_{4}}. \label{magact}
\end{equation}
One can compute the action for the electric instanton by looking at
the fixed point sets of $\partial_{\phi}$, and treating the field as 
``magnetic'' with respect to this isometry; we then find that the
action for the electric instanton is $-2\pi/A G_{4}$. 

This gives the correct actions for the compact solutions; to describe
the Lorentzian evolution, we choose the boundary which
subdivides the instanton to be the hypersurface 
$\psi = 0$, $\psi = \pi$, where the coordinate $\psi$ parametrises
the imaginary time. Since the division of the manifold $M$ must divide the
action symmetrically, and both fixed point
sets of $\partial_{\psi}$ in the original manifold contribute equally,
one must include half of each fixed point set in each of $M_{+}$ and $M_{-}$.

For the magnetic solution, the action is half of
(\ref{magact}), and thus the total entropy for the process 
is given by $2\pi/B$. For the
electric solution, we must add a boundary Maxwell field term, and the
entropy is given by the same expression. The pair creation rates,
which are given by the exponentiation of the entropy, are identical 
in the two cases. Note that our actions and entropies
are in agreement with those calculated in \cite{HR}, \cite{MaRo}. 

\bigskip

In discussing higher dimensional generalisations we run into the
problem that few compact solutions of such theories have been
constructed. We therefore discuss here
the simplest generalisation to a limit of the
five dimensional Reissner-Nordstr\"{o}m de Sitter solution 
which is given by 
\begin{equation}
ds^2 = \frac{1}{A} (d\chi^2 + \sin^2\chi d\psi^2) + \frac{2}{B}
d\Omega_{3}^2,
\end{equation}
with the latter term being the standard metric on a unit three
sphere. Again we work within Einstein-Maxwell theory, since the
dilaton field is unimportant here. 

It is straightforward to determine the
cosmological constant as $\Lambda = (A+2B)/3$ 
where the electric field is given by
\begin{equation}
F = -i q (\frac{B^{3/2}}{2\sqrt{2}A}) \sin\chi d\chi \wedge d\psi,
\end{equation}
with the charge being given by 
\begin{equation}
q^2 = \frac{(B-A)\sqrt{2}}{A B^{3/2}}.
\end{equation}
Evidently there is no corresponding magnetic solution. 
As in the four dimensional solution, the coordinate $\chi$ runs from
$0$ to $\pi$, and $\psi$ will parametrise the imaginary time. One can
then verify explicitly that the action for the compact manifold is
given by the fixed point set expression, where we take the Killing
vector to be $\partial_{\psi}$ which has fixed three spheres at the
poles. 

The Lorentzian evolution is again described by taking the boundary surface
$\psi = 0, \pi$, and choosing half of each fixed point set to lie in
each half of the manifold. The entropy of the solution will then be
given by one quarter of the three volume of the fixed point set at 
$\chi=0$, which we interpret as a cosmological horizon, and one
quarter of that of the three sphere $\chi = \pi$, which we interpret
as a black hole horizon. 
Thus our results demonstrate not only the well-known result that
cosmological and black hole horizons have an intrinsic entropy equal
to one quarter of their areas, but also extend the proof to gauge
field theories and higher dimensions. 

Our treatment of $(p+1)$ form gauge fields also
demonstrates that pair creation of $p$-branes is associated
with an entropy equal to the volumes of the horizons.
For electric fields we will again need to include a boundary term to
ensure that we take variations over solutions with constant charges. 
The boundary term that we require is 
\begin{equation}
S_{boundary} = - \frac{(p+1)}{8\pi G_{d}} \int_{\Sigma_{i}} d^{d-1}x
e^{-a\Phi}{\cal H}^{\mu_{1}....\mu_{p+1}} n_{\mu_{1}} {\cal
  B}_{\mu_{2}....\mu_{p+1}},
\end{equation}
where $n$ is the normal to the boundary, and ${\mathcal{B}}_{p}$ is a $p$ form
potential such that ${\mathcal H} = d {\mathcal B}_{p}$. It is
straightforward to verify that such a choice of boundary term gives
the required variational behaviour. As before
this boundary term precisely cancels out the gauge field volume term,
and thus the entropy for the process is given by the sum over
fixed point sets. 

\section{Extreme solutions} \label{exsol}
\noindent

We now consider the treatment of extreme solutions, which in this context
means compact Euclidean solutions with internal boundaries to the manifold. 
The action for a
solution of the generic theory discussed in the previous section will then 
have an additional term deriving from the extrinsic curvature of the boundary
$\partial M$ of the $d$ dimensional manifold $M$
\begin{equation}
S_{boun} = - \frac{1}{8 \pi G_{d}} \int_{\partial M} d^{d-1}x 
{\cal K} \sqrt{b}. \label{boin}
\end{equation}
As before one can decompose the volume term in terms of the dilation current; 
for a magnetic solution the total action for the Euclidean solution then
becomes
\begin{equation}
S_{E} = - \frac{\beta}{8 \pi G_{d}} \int_{\Sigma} dJ_{D} + S_{boun}.
\label{bugs}
\end{equation}
The boundary of the $(d-1)$ dimensional hypersurface now includes 
contributions not only from the fixed point sets of the imaginary time 
Killing vector, but also contributions from the dimensional 
reduction of the original boundary $\partial M$. For a pure electric field we 
again obtain the additional term from a volume integral of the field
\begin{equation}
\frac{1}{8\pi G_{d}} \int_{M} d^dx \sqrt{\hat{g}} e^{- a\Phi} \bar{F}^2.
\end{equation}
In the context of the no boundary proposal this term cancels with that
on the initial value hypersurface, and the total action for both electric
and magnetic pair creation processes is given by one half of (\ref{bugs}), 
where only $(d-2)$ dimensional fixed point sets are possible.

Now one cannot relate the boundary geometry term to the gauge fields
without further assumptions about the topology.
We shall discuss two generic types of solution with internal boundaries that
are physically interesting, both of which in some sense 
represent pair creation of extreme black holes \footnote{For
  discussions of the physical interpretations of all these pair
  creation solutions see \cite{MaRo}.}. 
The first, usually referred to as ``cold'' cosmological
pair creation, is a solution of topology $R^2 \times S^{d-2}$, which has a 
boundary of topology $S^1 \times S^{d-2}$ and a fixed point set of the 
imaginary time Killing vector within the manifold. The form of the metric
is
\begin{equation}
ds^2 = dR^2 + R^2 d\psi^2 + \alpha d\Omega_{d-2}^2,
\end{equation}
where $\psi$ is the imaginary time, $\alpha$ is related to the 
cosmological constant and the coordinate $R$ runs from the origin $R=0$
to $R_{\infty}$ which we can take to infinity at the end of the
calculation. One can choose the other fields so that the solution
satisfies the field equations. 
The metric is globally a product, and the extrinsic curvature of the 
boundary is non trivial, so that the boundary term (\ref{boin}) is given by
minus one quarter of the volume of the $(d-2)$ sphere. The volume term in
(\ref{bugs}) can be decomposed as
\begin{equation}
S_{E} = \frac{V_{d-2}(R_{\infty})}{4G_{d}} - \frac{V_{d-2}(R=0)}{4G_{d}},
\end{equation}
where we have taken account of the direction of the normal to the boundary at 
infinity. 
Note that the term at
infinity is obtained by integrating the dilation current over the
boundary; the form of the metric then implies that it reduces to the volume. 
Thence the boundary extrinsic curvature term cancels with the
term from the dilation current, and the total entropy for the pair creation
process is given by
\begin{equation}
S = \frac{V_{d-2}(R=0)}{4G_{d}},
\end{equation}
that is, by the volume term for the horizon contained within the Euclidean 
solution.

The second type of solution, referred to as ``ultracold'' cosmological pair 
creation, is a solution of topology $R^2 \times S^{d-2}$, which has a boundary
of topology $R^1 \times S^{d-2}$ and no fixed point sets of the imaginary 
time Killing vector within the manifold. The form of the metric is 
\begin{equation}
ds^2 = dx^2 + dy^2 + \alpha_{1} d\Omega_{d-2}^2,
\end{equation}
where $y$ is the imaginary time, $\alpha_{1}$ is related to the cosmological
constant and the coordinate $x$ runs from $-x_{\infty}$ to $x_{\infty}$, 
where we can take $x_{\infty}$ to infinity at the end of the calculation. 
The extrinsic curvatures of the boundaries at $\pm x_{\infty}$ vanish, and 
hence there is no boundary term in the action. The volume term in (\ref{bugs})
can be decomposed in terms of the integral of the dilation current over
only the $(d-2)$ volumes of the boundaries, since
there are no fixed point sets of the Killing vector, where we take account
correctly of the directions of the normals.
Since the dilation current vanishes throughout the manifold, the
total action for the solution vanishes, as does the entropy of
the pair creation process, as we would expect, since there are no
horizons contained within the manifold. Our results are in agreement
with those given in \cite{MaRo} for four dimensional solutions. 

\section{Conclusions}
\noindent

In this paper, we have discussed the action of a circle isometry group
on compact Euclidean Einstein manifolds, and on compact Euclidean
solutions of supergravity theories. For the former, we
can decompose the action of the solution in terms of characteristic
properties of the action of any Killing vector with closed orbits. 
Although in four dimensions the characteristic properties of the fixed
point sets, that is the volume and nut behaviour, alone determine the
action, in higher dimensions one has to take account also of
non-trivial $(d-3)$ cohomology of the orbit space. It is natural to
characterise the contribution from the cohomology as an integral over
the entire orbit space rather than from individual fixed point sets. 
Introducing a microcanonical ensemble, the entropy is equal to minus
the action, and thence we obtain an expression for the entropy in
terms of the volume of $(d-2)$ dimensional fixed point sets, and the
cohomology of the orbit space. 

\bigskip

In the context of the no boundary proposal, our decomposition of the
action allows us to calculate the entropy of a particular solution, and
thence the probability of a process occurring, in terms of these fixed
point sets. In particular, we can demonstrate explicitly that pair
creation of (neutral) black p-branes in a cosmological background
is associated with an entropy
equal to one quarter of the volumes of the horizons. 

It is natural to extend the analysis to compact Euclidean solutions of
Einstein gravity, coupled to scalar and gauge fields; with suitable
couplings between the fields this includes solutions
of supergravity theories. We
find that the action can again be decomposed in terms of
characteristics of the action of an isometry
but that for electric fields there is an additional term
left over. In the context of the no boundary proposal, this
additional term can be regarded as a constraint on the initial value
hypersurface that we fix the electric charge (per unit area). Thus
we are able to demonstrate that generic black $p$-branes pair created in an 
appropriate background have an entropy equal to one quarter of the
volumes of the event and cosmological horizons. 
We also considered briefly the extremal limits of pair creation
solution, for which we must consider internal boundaries to the
manifold. 

\bigskip

The analysis of non-compact Euclidean Einstein manifolds is more subtle. 
Firstly, we will require the existence of a suitable background with 
respect to which all thermodynamic quantities can be defined. Secondly,
we have to identify an appropriate temperature before we can define
the entropy; we must work within a canonical or grand canonical
ensemble. When we attempt to decompose the action in terms of the
the action of an isometry in general there will
be surface terms on the boundary at infinity which are left
over. If we decompose the action in terms of the action of a Killing
vector which has null fixed point sets in the Lorentzian continuation,
these surface terms will be related to the energy and angular
momentum, and the sum over the fixed point sets can again be
identified as the entropy. A discussion of the action of isometries on
non compact manifolds is contained in \cite{MMT}.

\end{document}